\documentclass[oldversion, letter]{aa} 
\usepackage{graphicx}
\usepackage{txfonts}

\begin{document}

\title{Fueling the central engine of radio galaxies}

\subtitle{I. The molecular/dusty disk of 4C~31.04\thanks{Based on observations carried out with the IRAM Plateau de Bure Interferometer. IRAM is supported by INSU/CNRS (France), MPG (Germany) and IGN (Spain).}}

\author{S.~Garc{\'{\i}}a-Burillo\inst{1}
  \and F.~Combes \inst{2}
     \and R.~Neri\inst{3}
     \and A.~Fuente\inst{1}
     \and A.~Usero\inst{1,4}
     \and S.~Leon\inst{5}
     \and J.~Lim\inst{6}}


\institute{Observatorio Astron{\'o}mico Nacional, Alfonso XII, 3, 28014 Madrid, Spain
\email{s.gburillo@oan.es, a.fuente@oan.es}
  \and 
	Observatoire de Paris, LERMA, 61, Av. de l'Observatoire, 75014 Paris, France
\email{francoise.combes@obspm.fr}
  \and Institut de Radioastronomie Millim\'etrique, 300, Rue de la Piscine, 38406 St Mt d'H\`eres, France
\email{neri@iram.fr}
  \and Centre for Astrophysics Research, University of Hertfordshire, College Lane, AL10 9AB, UK
\email{a.usero@herts.ac.uk}
   \and Instituto de Astrof{\'{\i}}sica de Andaluc{\'{\i}}a, C$^{o}$ Bajo de Hu\'etor, 50, 18008 Granada, Spain
\email{stephane@iaa.es}
   \and Inst. of Astron. and Astrophysics, Academia Sinica, P.O. Box 23-141, Taipei 106, Taiwan 
\email{jlim@asiaa.sinica.edu.tw}
}

\date{Received  / Accepted }

\abstract{We report the detection of a massive (M$_{gas}$$>$5$\times$10$^{9}$M$_{\sun}$) molecular/dusty disk of 1.4\,kpc-size fueling the central engine of the Compact Symmetric Object (CSO) 4C\,31.04 based on high-resolution (0.5$\arcsec$--1.2$\arcsec$) observations done with the IRAM Plateau de Bure interferometer (PdBI). These observations allow for the first time to detect and map the continuum emission from dust at 218\,GHz in the disk of a CSO. The case for a massive disk is confirmed by the detection of strong HCO$^{+}$(1--0) line emission and absorption. The molecular gas mass of 4C\,31.04 is in the range 0.5$\times$10$^{10}$--5$\times$10$^{10}$M$_{\odot}$. While the distribution and kinematics of the gas correspond roughly to those of a rotating disk, we find evidence of distortions and non-circular motions suggesting that the disk is not in a dynamically relaxed state. We discuss the implications of these results for the general understanding of the evolution of radio galaxies.}

\maketitle

\section{Introduction}

%
Contrary to the well developed radio jets of classical radio galaxies (on scales of $\sim$a few~10--100\,kpc), the distinguishing property of Compact Symmetric Objects (CSOs) is the small separation between their radio-lobes ($\sim$a few\,100pc--1\,kpc). Two scenarios were proposed to account for the radio structure of CSOs: the {\it youth} scenario, which contemplates CSOs as young radio galaxies with jets in the making (e.g. Fanti et al.~\cite{fan95}, Owsianik \& Conway~\cite{ows98}), and the {\it frustration} scenario, which interpreted the absence of large jets as a consequence of the confinement by a highly dense massive cocoon surrounding the central engine (e.g. O' Dea et al.~\cite{ode91}, Bicknell et al.~\cite{bic97}). 
Of particular note, there is mounting evidence that interaction between the radio plasma and the ISM present around the central engines of radio-loud galaxies is at work also in CSOs (Morganti et al.~\cite{mor04}) .

A study of the content, distribution and kinematics of cold gas via molecular and atomic lines is key to elucidating the nature of CSOs and their evolutionary link to other radio-loud galaxies. 4C\,31.04 is a nearby CSO at z$\sim$0.06 at the nucleus of the giant elliptical MCG 5-8-18, located at the center of a cluster. HI absorption was detected by Van Gorkom et al.~(\cite{van89}) and Mirabel~(\cite{mir90}), and mapped at high-resolution by Conway~(\cite{con96}). The HST color maps obtained by Perlman et al.~(\cite{per01})(Pe01) unveil that the optical image of 4C\,31.04 is permeated with obscuration features. These consist of an elongated $\sim$1''--central disk oriented along PA$\sim$5$^{\circ}$-10$^{\circ}$, surrounded by a spiral-like structure which stretches from the disk over the central $\sim$4$\arcsec$ of the image. The central disk is roughly perpendicular to the axis of the double-lobe radio source shown in the 5\,GHz VLBA image of Giovannini et al.~(\cite{gio01})(Gi01). This radio source consists of two lobes that are separated by $\sim$0.07$\arcsec$(80\,pc) and extend close to the plane of the sky.

We have probed the gas and dust content of 4C\,31.04, by observing the HCO$^+$(1--0) and $^{12}$CO(2--1) molecular lines and their underlying continuum emission with unprecedented resolution ($\leq$1$\arcsec$) using the IRAM Plateau de Bure Interferometer (PdBI). In this Letter we discuss the results obtained from the analysis of the PdBI maps of continuum emission detected at 1mm and 3mm. We also analyze the HCO$^+$(1--0) map of the galaxy showing emission and absorption of this line. The CO(2--1) molecular line has only been detected towards the central offset of the galaxy due to insufficient sensitivity. The results obtained show the existence of a highly disturbed massive (M$_{gas}$$>$5$\times$10$^{9}$M$_{\sun}$) molecular disk fueling the central engine of 4C\,31.04. 

   \begin{figure}
   \centering
   \includegraphics[width=8.5cm]{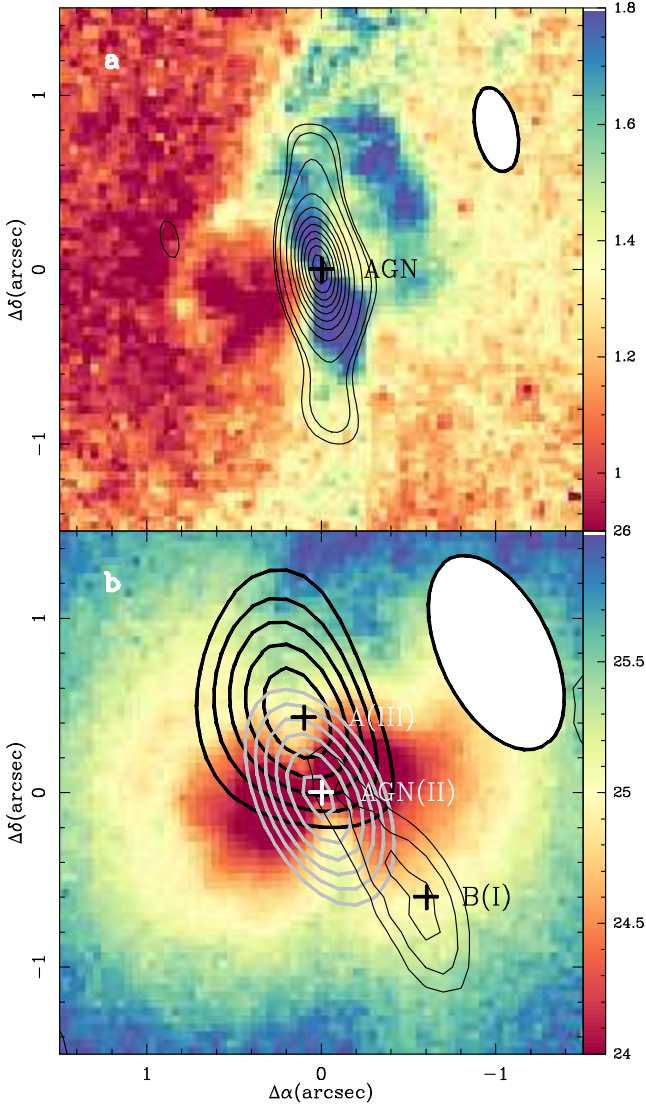}
   \caption{{\bf a)} The 1mm-continuum PdBI map (at 218\,GHz) of 4C\,31.04 (contours: 3$\sigma$, 4$\sigma$, 7$\sigma$ to 55$\sigma$ in steps of 6$\sigma$; 1$\sigma$=0.8\,mJy\,beam$^{-1}$) is overlaid on the R-H color image of Pe01 from HST (color scale). ($\Delta\alpha$, $\Delta\delta$)--offsets in arcsec are relative to the location of the AGN. See discussion in Sect.~\ref{dust-emi} for details.  
{\bf b)} The HCO$^{+}$\,(1--0) line maps of 4C\,31.04 obtained after subtraction of the continuum source for the three velocity channels defined in Fig.\,\ref{Fig3}: {\bf I}(emission)=[-300,25]\,km\,s$^{-1}$(black thin contours; with maximum at {\bf B}), {\bf II}(absorption)=[25,275]\,km\,s$^{-1}$(grey contours; with maximum at the AGN), {\bf III}(emission)=[275,900]\,km\,s$^{-1}$ (black thick contours; with maximum at {\bf A}). Levels are 3.5$\sigma$, 4$\sigma$, and 4.5$\sigma$ for channel {\bf I} (1$\sigma$({\bf I})=0.06Jy\,km\,s$^{-1}$), -40$\sigma$, to -100$\sigma$ in steps of -10$\sigma$ for channel {\bf II} (1$\sigma$({\bf II})=0.05Jy\,km\,s$^{-1}$) and 5$\sigma$ to 13$\sigma$ in steps of 2$\sigma$ for channel {\bf III} (1$\sigma$({\bf III})=0.08Jy\,km\,s$^{-1}$). HCO$^{+}$ contours are overlaid on the F702W HST image of Pe01 (color scale in mag.pixel$^{-1}$). Ellipses represent the beams at 218\,GHz({\bf a)}) and 84\,GHz({\bf b)}). See discussion in Sect.~\ref{mol-emi} for details.}
         \label{Fig1}
   \end{figure}
%

\section{Observations}
We observed 4C\,31.04 simultaneously in the
HCO$^+$(1--0) and $^{12}$CO(2--1) transitions
with the PdBI in the A+ configuration on February 2006. 
We used one receiver setting centered on the HCO$^+$ transition (redshifted to 84.204\,GHz) and two overlapping settings (centered each at 217.433\,GHz and 217.873\,GHz) to sample the 
$^{12}$CO(2--1) line (redshifted to 217.653\,GHz). These receiver settings translate into a
velocity coverage of $\sim$2000km\,s$^{-1}$ at 84\,GHz and
$\sim$1400km\,s$^{-1}$ at 218\,GHz. The phase tracking center was set
at ($\alpha_{2000}$, $\delta_{2000}$)=(01$^{\rm h}19^{\rm m}35.001^{\rm s}$,
32$^\circ$10$\arcmin$50.06$\arcsec$) coincident with
the position of the radiocontinuum source core (Gi01). 
The line visibilities at 84\,GHz were self-calibrated on the point-like radio-continuum 
source in the 84.402$-$84.475\,GHz band.  
The 1$\sigma$-noise for an on-source integration time of 9\,hrs is 0.6\,mJy\,beam$^{-1}$ in 
7.5\,MHz-channels.  At 218\,GHz the visibilities were calibrated using the antenna
based scheme.  The point-source sensitivity at 218\,GHz is 
1.9\,mJy\,beam$^{-1}$ in 15\,MHz-channels. Maps were obtained with uniform
weighting and yielded synthesized beams of 0.50$\arcsec
\times$0.24$\arcsec$\,@PA=12$^\circ$ at 218\,GHz and 1.20$\arcsec
\times$0.64$\arcsec$\,@PA=26$^\circ$ at 84\,GHz. 
We have obtained a map of the continuum emission at 218\,GHz in 4C\,31.04 using the 1GHz-wide image side band of the 1mm receivers. Inside this window, there is no contribution from line emission. Similarly, a 3mm continuum map (at 84\,GHz) has been built using velocity channels free of line emission in the signal side band of the PdBI receivers. The corresponding 1-$\sigma$ sensitivities are $\sim$0.8\,mJy\,beam$^{-1}$ and $\sim$0.5\,mJy\,beam$^{-1}$ at 218\,GHz and 84\,GHz, respectively. Although re-determined in this work, the velocity scale is initially referred to the redshift z$=$0.0592 derived by Sargent~(\cite{sar73}). Luminosity and angular distances are D$_{L}$=260\,Mpc and D$_{A}$=230\,Mpc; the latter gives 1$\arcsec$=1.1\,kpc.

\section{Results}
\subsection{Continuum maps: detection of a dusty disk}\label{dust-emi}

 The 1mm map shown in Fig.\,\ref{Fig1}a reveals a strong emission source in the nucleus of 4C\,31.04. The source can be decomposed into an unresolved central component, and a disk.
The disk extends over $\sim$1.6$\arcsec$($\sim$1.8\,kpc) along PA$\sim$5$^\circ$-10$^\circ$. The measured 1mm-continuum visibilities can be fitted by a $\sim$40\,mJy point source centered at the AGN, and an elliptical Gaussian source  which has a deconvolved FWHM-size of 1$\arcsec \times$0.1$\arcsec$ (1.1\,kpc$\times$100\,pc) and an integrated flux of $\sim$20\,mJy. The continuum emission at 3mm consists of a point source of $\sim$160\,mJy with no extended disk detected to the limits of our spatial resolution and sensitivity. The point source detected at 1mm and 3mm is the higher frequency counterpart of the non-thermal double-lobe radio source mapped at cm-wavelengths in this CSO (e.g., Gi01). The spectral index ($\alpha$) of the total continuum emission between 84\,GHz and 218\,GHz (with S$_{\nu}\sim\nu^{-\alpha}$), confirms the dominance of the non-thermal central component: $\alpha$$\geq$1. 

The detected 1mm disk is tightly linked to the dusty disk identified in the  R-H color image of Pe01 (see Fig.\,\ref{Fig1}a). 
The dusty disk seems to be mostly edge-on and oriented perpendicular to the axis of the double-lobe radio source. However, thermal free-free emission from an embedded star forming episode could also partly contribute to the disk emission at 1mm. While there is no evidence of such a disk at cm wavelengths, it should be noted that the published VLBA/VLBI maps of 4C\,31.04 would not be able to recover any emission on the typical scales of the disk. However the non-detection of an extended disk at 3mm argues against a significant free-free emission component at 1mm. Furthermore, there is no detection of a thermal outflow-like component either at 1mm or at 3mm.

If the 1mm flux of the disk can be attributed to dust emission, a mass budget (M$_{dust}$) can be derived provided that values for the temperature (T$_{dust}$) and for the emissivity of dust ($\kappa$) are assumed. IRAS observations reveal significant amounts of warm dust in 4C\,31.04. From the flux ratio measured by IRAS S$_{\nu}$(100$\mu$)/S$_{\nu}$(60$\mu$)$\sim$3.5, we derive a range for T$_{dust}$=25--30\,K for a corresponding range of $\beta$=2--1 in the wavelength dependency of $\kappa \sim \lambda^{-\beta}$.
Taking $\kappa$(850$\mu$)=0.9\,cm$^{2}$gr$^{-1}$ from Klaas et al.~(\cite{kla01}), and T$_{dust}$=30\,K (i.e., the solution for $\beta$=1) we derive M$_{dust}\sim$6$\times$10$^{7}$M$_{\sun}$ for the warm dust. To better estimate the total dust mass we have fitted the fluxes measured by IRAS (at 100$\mu$ and 60$\mu$) and the PdBI (at 1.3mm) using a two component model for the dust temperatures. Any plausible fit to the observations gives a value for M$_{dust}$ of (4--7)$\times$10$^{8}$M$_{\sun}$. The higher (lower) value of M$_{dust}$ corresponds to the fit with $\beta$=1.5 ($\beta$=1) and temperatures of $\sim$16\,K(21\,K) and $\sim$32\,K(39\,K) for the cold and warm dust components, respectively. The cold dust component seems to contain the bulk of the mass: M$_{dust}^{cold}>$0.95$\times$M$_{dust}^{total}$).

Taking a molecular gas--to--dust ratio $\sim$100 (Seaquist et al.~\cite{sea04}), we foresee that the total molecular gas mass of the disk might be M$_{gas}$$\sim$(4--7)$\times$10$^{10}$M$_{\sun}$. If we allow for a minor contribution of free-free emission to the 1mm flux, this can be taken as an upper limit to M$_{gas}$. As argued in Sect.~\ref{mol-emi}, the case for a massive gas reservoir in 4C\,31.04 is nevertheless supported by the detection of HCO$^{+}$(1--0) emission from a rotating molecular disk.

   \begin{figure}
   \centering
   \includegraphics[width=8.5cm]{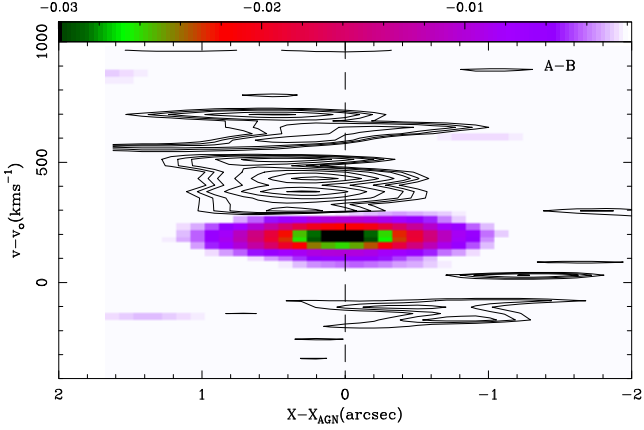}
      \caption{The kinematics of molecular gas along the strip connecting points {\bf A} and {\bf B} (emission peaks for HCO$^+$(1--0)) are illustrated by a position-velocity diagram. Contour levels represent emission of HCO$^+$(1--0): 2$\sigma$, 2.5$\sigma$, 3$\sigma$ to 7$\sigma$ in steps of 1$\sigma$; 1$\sigma$=0.6\,mJy\,beam$^{-1}$. Color scale (mJy\,beam$^{-1}$) represents absorption in HCO$^+$(1--0) near the AGN. X-offsets are in arcsec relative to the position closest to the AGN. Velocity scale in Y axis referred to v$_o$(z=0.0592).} 
         \label{Fig2}
   \end{figure}
%

\subsection{Line emission: detection of a molecular disk}\label{mol-emi}

\subsubsection{Distribution and kinematics}

\begin{figure}
   \centering
   \includegraphics[width=8.5cm]{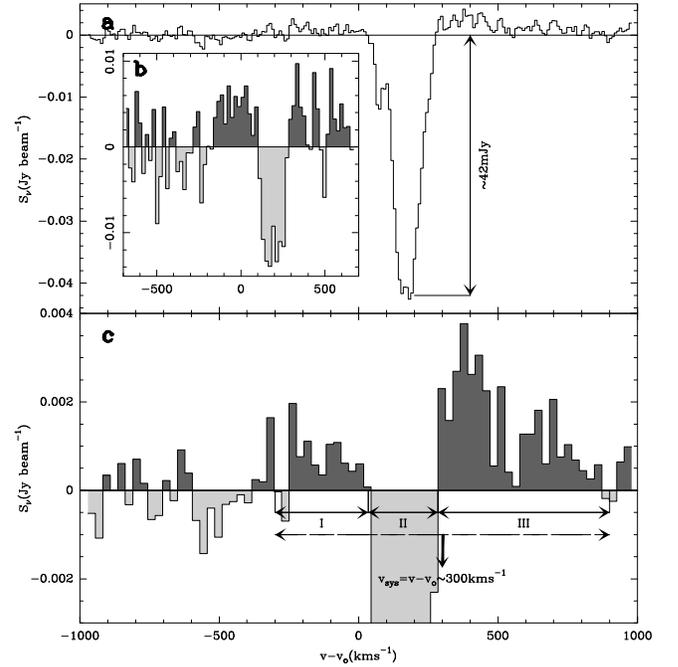}
      \caption{{\bf a)} Spectrum of the HCO$^{+}$(1--0) line emission and absorption at the central offset (AGN) of 4C\,31.04. A strong absorption line is detected against the 3mm continuum source. Velocity scale is referred to v$_{o}$, which corresponds to z=0.0592. {\bf b)} Same as {\bf a)} but for the CO(2--1) line. {\bf c)} A zoomed view on the spectrum shown in {\bf a)} helps to delimit intervals {\bf I}--to--{\bf III} and derive a new value for v$_{sys}$.}
         \label{Fig3}
   \end{figure}

The channel maps obtained after subtraction of the 3mm continuum source show that HCO$^{+}$(1--0) is detected both in emission and absorption in 4C\,31.04. As shown in Fig.\,\ref{Fig1}b, HCO$^{+}$(1--0) is detected in emission in two velocity ranges: from --300 to 25\,km\,s$^{-1}$, and from 275 to 900\,km\,s$^{-1}$ (channels {\bf I} and {\bf III}, respectively; see Fig.\,\ref{Fig3}). The strongest emission ({\bf III}) peaks NE with respect to the AGN (offset A=[0.1$\arcsec$, 0.4$\arcsec$] in Fig.\,\ref{Fig1}b). Compared to {\bf III}, emission in {\bf I} is $\geq$3 times weaker, and peaks SW with respect to the AGN (offset B=[--0.6$\arcsec$, --0.6$\arcsec$] in Fig.\,\ref{Fig1}b). In  contrast, the line is detected in absorption from 25 to 275\,km\,s$^{-1}$ (channel {\bf II}; see Fig.\,\ref{Fig3}). The HCO$^+$ absorption is detected as a point source centered at the AGN. These results indicate that the distribution of molecular gas traced by HCO$^+$ has been spatially resolved by the PdBI. 

At first order, the emission and absorption of HCO$^+$ seem to arise from a rotating disk of $\sim$1.4\,kpc-size. This is illustrated in Fig.\,\ref{Fig2} which shows the position-velocity (p-v) plot along the strip connecting A and B, i.e., close to the apparent kinematic major axis of the disk. By assuming that the total width (FWZP) of the HCO$^+$ emission profile is twice the projected rotational velocity ($v_{rot}$) we derive $v_{rot} \sim$500kms$^{-1}$ for an edge-on disk geometry and a value for the systemic velocity (v$_{sys}$) which is +300\,km\,s$^{-1}$-redshifted with respect to the redshift initially assumed in this work (z=0.0592). The HCO$^+$-based redshift (z(HCO$^+$)=0.0602$\pm$0.0002) agrees with the values recently derived from optical spectroscopy observations of 4C\,31.04 (z=0.0600$\pm$0.0001; Marcha et al.~\cite{mar96}).

If circular motions completely ruled the gas kinematics, the absorption profile against the AGN (this being the dynamical center) should be centered at v$_{sys}$. This is however at odds with observations: as shown in Fig.\,\ref{Fig3}, the gas seen in absorption in channel {\bf II} is $\sim$150\,km\,s$^{-1}$ blue-shifted on average with respect to z=0.0602 (similarly to the HI absorption of Conway~\cite{con96}). This result can only be reconciled if the gas causing the absorption is outflowing from the nucleus or subject to strong non-circular motions, thus partly invalidating the initial premises. An alternative solution adopting the center of the absorption profile as v$_{sys}$ would only enhance the asymmetry in the HCO$^+$ emission profile. Furthermore, there is evidence of distortions in the molecular disk from the observed distribution. The SW side of the molecular disk is offset from the dust disk: the HCO$^+$ disk is tilted towards a larger PA (Fig.\,\ref{Fig1}b). A conspiracy of absorption and emission may explain the observed anticorrelation. However, independent evidence of distortions is also found on the opposite side of the disk:  whereas the bulk of the HCO$^+$ emission NE follows the dust disk along PA=5$^{\circ}$-10$^{\circ}$, the molecular disk is seen to thicken along its minor axis in this region (Fig.\,\ref{Fig1}b).

While the reasons explaining the observed distortions in the distribution and kinematics of the gas cannot be elucidated with these observations, these results further illustrate that the rotating molecular disk of 4C\,31.04 is not dynamically relaxed.

%

\subsubsection{Masses}

The HCO$^{+}$ line in emission probes the dense phase (n(H$_2$)$>$10$^{4}$cm$^{-3}$) of the molecular disk of 4C\,31.04. Without further constraints at hand, an estimate of the mass of dense molecular gas (M$_{dense}$) can be derived from the luminosity of the line (L[HCO$^{+}$]=4.2$\times$10$^8$K\,km\,s$^{-1}$\,pc$^{2}$) if we assume low opacities, LTE conditions, an abundance for HCO$^+$ (X[HCO$^+$]), and an excitation temperature (T$_{ex}$) for the line. For T$_{ex}$=10\,K and X[HCO$^+$]=10$^{-8}$, we derive M$_{dense}\sim$5$\times$10$^8$M$_{\odot}$. This is probably a lower limit, because the HCO$^{+}$ line is likely optically thick (see Fig.~\ref{Fig3}). Alternatively, we can derive  M$_{dense}$ via a L(HCO$^{+}$)--to--M$_{dense}$ conversion factor if we assume that the emission comes from a clumpy medium of optically thick virialized clouds. The value of the conversion factor would depend on the physical parameters of the clouds through the ratio $\sim$n$^{1/2}$(H$_2$)/T$_b$(HCO$^+$), where n(H$_2$) is the molecular gas density and T$_b$(HCO$^+$) is the emergent line brightness temperature of the clouds. Through LVG simulations, assuming a range for T$_K$=20--to--60\,K and n(H$_2$)=10$^{4}$--to--10$^{5}$cm$^{-3}$, we derive an average conversion factor M$_{dense}$/L(HCO$^{+}$)$\sim$10\,M$_{\odot}$\,(K\,km\,s$^{-1}$\,pc$^{2}$)$^{-1}$.
This would imply that M$_{dense}$$\sim$4$\times$10$^9$M$_{\odot}$. Taking HCO$^+$(1--0)/CO(1--0)=0.05 (Graci\'a-Carpio et al.~\cite{gra06}) we derive M$_{gas}$/M$_{dense}$$\sim$10, and thus estimate M$_{gas}$ to range between $\sim$5$\times$10$^9$M$_{\odot}$ and $\sim$4$\times$10$^{10}$M$_{\odot}$.

The sensitivity reached in CO(2--1) was insufficient to map the emission of this line in 4C\,31.04, therefore M$_{gas}$ cannot be derived from these data. The line is detected in absorption and emission only towards the AGN (Fig.~\ref{Fig3}). However, Oca\~na-Flaquer et al.(in prep) recently detected in 4C\,31.04 the emission and absorption of CO(1--0) and CO(2--1) using the IRAM 30m telescope, improving previous searches for CO emission (O'Dea et al.~\cite{ode05}; Evans et al.~\cite{eva05}). Using a CO--to--H$_2$ conversion factor N(H$_2$)/I$_{CO}$=2.2$\times$10$^{20}$cm$^{-2}$\,K$^{-1}$\,km$^{-1}$\,s (Solomon \& Barrett~\cite{sol91}), Oca\~na-Flaquer et al. estimate M$_{gas}$$\sim$5$\times$10$^9$M$_{\odot}$, a value within the range derived from HCO$^+$. Note that due to the superposition of absorption and emission in the profiles, the values derived from molecular line emission underestimate M$_{gas}$.

Taken the estimates from dust continuum and line emission together, we conclude that M$_{gas}$ lies in the range (0.5--5)$\times$10$^{10}$M$_{\odot}$. This is much larger than the mass of neutral hydrogen derived by Conway~(\cite{con96}) from HI line data ($\sim$10$^8$M$_{\odot}$)  

\subsection{Line absorption}\label{mol-abs}
Assuming LTE conditions we can derive the column densities of HCO$^+$(N$_{HCO^+}$) and CO(N$_{CO}$) from the the J$\rightarrow$J+1 absorption lines detected towards the center of 4C\,31.04 according to:

\[N_{mol} = \frac{8 \pi \nu^3}{c^3 A_{ul} g_u} \frac{ Q(T_{\rm ex})
 \exp{(E_J/k T_{\rm ex})}} {(1-\exp{(-h \nu / k T_{\rm ex})})} \int \tau dV \] 

\noindent where E$_J$ is the energy of the lower level of the transition, $T_{\rm ex}$ is the excitation temperature,  
Q(T$_{\rm ex}$) = $\sum_{Ji} (2Ji+1)\exp{(-E_{Ji}/kT_{\rm ex})}$ the partition
function and A$_{ul}$ the Einstein coefficient of the line.
A lower limit to the velocity integrated optical depth of the line $\int \tau dV$, and thus to $N_{mol}$, can be derived assuming that the covering factor $f$ of the absorbing molecular gas with respect to the extent of the radio source in 4C\,31.04 is unity. In this case we replace $\int\tau dV$ by $\tau_{obs}\Delta v$, where $\tau_{obs}$=ln(T$_{mb}^{C}$/T$_{mb}^{L}$) with T$_{mb}^{C}$ and  T$_{mb}^{L}$ being the main brightness temperature of the continuum and the line, respectively, and $\Delta v$ the Gaussian width of the line. From $\tau_{obs}$ we estimate a lower limit to $f$ which is similar for the two lines ($f>$1-e$^{-\tau_{obs}}\sim$0.27), suggestive of a common value for $f$. Assuming $f$=1 and $T_{\rm ex}$=10\,K for the two lines, we derive N(HCO$^+$)/N(CO)=1.6$\times$10$^{-3}$. While this value is surprisingly high relative to galactic standards  (see Table~3 of Wiklind \& Combes~\cite{wik95}, [Wi95]), it likely overestimates the true HCO$^+$/CO abundance ratio for two reasons. First, $T_{\rm ex}$ is probably smaller for the HCO$^+$ line than for the CO line. Second, the chances for the CO absorption to be underestimated due to corruption by emission within the beam are much higher than for the HCO$^+$ absorption  (see discussion in Wi95).


\section{Conclusions} 

The high-resolution and high-sensitivity capabilities of the PdBI have allowed for the first time to detect and map a massive molecular/dusty disk fueling the central engine of a CSO like 4C\,31.04. 
Based on the estimates derived from the continuum and line emissions, the molecular disk mass M$_{gas}$ lies in the range (0.5--5)$\times$10$^{10}$M$_{\odot}$. Even if we adopt the lower bound to M$_{gas}$, these observations reveal the existence of significant amounts of molecular gas in the central 1.4\,kpc disk of 4C\,31.04, comparable to those identified in ULIRGs ($\sim$10$^9$--5$\times$10$^{10}$M$_{\odot}$; Sanders et al.~\cite{san91}), and much larger than the masses of cold gas detected in normal elliptical galaxies ($\sim$10$^7$--10$^{8}$M$_{\odot}$; Lees et al.~\cite{lee91}). Of particular note, the large value of M$_{gas}$ derived for 4C\,31.04 would be roughly consistent with the scenario of confinement in this radio source. Using simple analytical models, Carvalho~(\cite{car98}) predicts that a clumpy medium with M$_{gas}$$\sim$10$^9$--10$^{10}$M$_{\odot}$ can confine a radio source on the scale of 0.5--1\,kpc. In the case of 4C\,31.04, the bulk of M$_{gas}$ is seen to be in a disk roughly perpendicular to the radio source axis. This particular geometry would decrease the efficiency of frustration, however.

 Within the frustration scheme, it is foreseen that a disturbed ISM around the radio AGN would reflect the interaction between the radio plasma and the dense confining cocoon. In the case of 4C\,31.04, there is evidence that a radio lobe-ISM interaction may be at work. The HST image of Pe01 shows cone-like features aligned with the radio axis of the source (see Fig.~\ref{Fig1}b), the likely signature of gas shocked by the jet (e.g., Labiano et al.~\cite{lab03}). Moreover, the distribution and kinematics of molecular gas probed by the HCO$^+$ observations presented in this work illustrate that the detected rotating disk is not in a fully relaxed state. The revealed distortions may reflect that the disk is still settling after a merger or an event of gas accretion. Alternatively, the jet and the cone-like features may be interacting with the disk and thus produce the reported distortions. 

Obtaining a better constrain to M$_{gas}$ in 4C\,31.04, and in CSOs in general, is key to exploring the nature of these sources. As illustrated in this work, the central gas disks of radio-galaxies are likely dominated by the molecular phase. In order to take full advantage of molecular line data, high spatial resolution and high sensitivity observations are needed to reduce at most the bias due to the confusion of absorption and emission in the beam. 




\begin{acknowledgements}
     We thank economic support from the Spanish MEC and Feder funds under grant ESP2003-04957 and from SEPCT/MEC under grant AYA2003-07584. We thank E.~Perlman for providing the HST data of 4C\,31-04.
\end{acknowledgements}


\begin{thebibliography}{}

\bibitem[1997]{bic97} Bicknell, G.~V., 
Dopita, M.~A., \& O'Dea, C.~P.~O.\ 1997, \apj, 485, 112 

\bibitem[1998]{car98} Carvalho, J.~C.\ 1998, \aap, 
329, 845 

\bibitem[1996]{con96} Conway, J.~E.\ 1996, IAU 
Symp.~175: Extragalactic Radio Sources, 175, 92 



\bibitem[2005]{eva05} Evans, A.~S., Mazzarella, 
J.~M., Surace, J.~A., Frayer, D.~T., Iwasawa, K., \& Sanders, D.~B.\ 2005, 
\apjs, 159, 197 



\bibitem[1995]{fan95} Fanti, C., Fanti, R., 
Dallacasa, D., Schilizzi, R.~T., Spencer, R.~E., \& Stanghellini, C.\ 1995, 
\aap, 302, 317 


\bibitem[2001]{gio01} Giovannini, G., 
Cotton, W.~D., Feretti, L., Lara, L., \& Venturi, T.\ 2001, \apj, 552, 508 [Gi01]

\bibitem[2006]{gra06} 
Graci{\'a}-Carpio, J., Garc{\'{\i}}a-Burillo, S., Planesas, P., \& Colina, 
L.\ 2006, \apjl, 640, L135 

 

\bibitem[2001]{kla01} Klaas, U., et al.\ 2001, 
\aap, 379, 823 

\bibitem[2003]{lab03} Labiano, A., et al.\ 
2003, Publications of the Astronomical Society of Australia, 20, 28 

\bibitem[1991]{lee91} Lees, J.~F., Knapp, G.~R., 
Rupen, M.~P., \& Phillips, T.~G.\ 1991, \apj, 379, 177

\bibitem[1996]{mar96} Marcha, M.~J.~M., 
Browne, I.~W.~A., Impey, C.~D., \& Smith, P.~S.\ 1996, \mnras, 281, 425

\bibitem[1990]{mir90} Mirabel, I.~F.\ 1990, \apjl, 
352, L37 

\bibitem[2004]{mor04} Morganti, R., 
Oosterloo, T.~A., Tadhunter, C.~N., Vermeulen, R., Pihlstr{\"o}m, Y.~M., 
van Moorsel, G., \& Wills, K.~A.\ 2004, \aap, 424, 119 


\bibitem[1991]{ode91} O'Dea, C.~P., Baum, 
S.~A., \& Stanghellini, C.\ 1991, \apj, 380, 66 

\bibitem[2005]{ode05} O'Dea, C.~P., Gallimore, 
J., Stanghellini, C., Baum, S.~A., \& Jackson, J.~M.\ 2005, \aj, 129, 610 

\bibitem[1998]{ows98} Owsianik, I., \& 
Conway, J.~E.\ 1998, \aap, 337, 69 

\bibitem[2001]{per01} Perlman, E.~S., Stocke, 
J.~T., Conway, J., \& Reynolds, C.\ 2001, \aj, 122, 536 [Pe01]


\bibitem[1991]{san91} Sanders, D.~B., 
Scoville, N.~Z., \& Soifer, B.~T.\ 1991, \apj, 370, 158 

\bibitem[1973]{sar73} Sargent, W.~L.~W.\ 1973, 
\apjl, 182, L13 

\bibitem[2004]{sea04} Seaquist, E., Yao, L., 
Dunne, L., \& Cameron, H.\ 2004, \mnras, 349, 1428 



\bibitem[1991]{sol91} Solomon, P.~M., \& 
Barrett, J.~W.\ 1991, IAU Symp.~146: Dynamics of Galaxies and Their 
Molecular Cloud Distributions, 146, 235 

\bibitem[1989]{van89} van Gorkom, J.~H., 
Knapp, G.~R., Ekers, R.~D., Ekers, D.~D., Laing, R.~A., \& Polk, K.~S.\ 
1989, \aj, 97, 708 

\bibitem[1995]{wik95} Wiklind, T., \& 
Combes, F.\ 1995, \aap, 299, 382 [Wi95]

\end{thebibliography}
\end{document}